\documentclass[12pt]{article}
\usepackage{amsmath,amssymb,epsfig,cite}
\usepackage{color}

\makeatletter

\@addtoreset{equation}{section} \makeatother

\newcommand{\be}{\begin{equation}}
\newcommand{\ee}{\end{equation}}
\newcommand{\beqa}{\begin{eqnarray}}
\newcommand{\eeqa}{\end{eqnarray}}

\begin{document}

\titlepage

\begin{flushright}
SU-4252-849 \\
\today \\
\end{flushright}
\vskip 1cm
\begin{center}
{\large \bf $Z^0 \rightarrow 2\gamma$ and the Twisted Coproduct of
the Poincar\'{e} Group }
\end{center}

\vspace*{5mm} \noindent

\centerline{A. P. Balachandran\footnote{bal@phy.syr.edu}}\vskip
0.2cm

\centerline{ \em Department of Physics, Syracuse University,
Syracuse, NY 13244-1130, USA.} \vskip 0.5cm

\vspace*{5mm} \noindent

\centerline{S. G. Jo\footnote{sgjo@knu.ac.kr}}\vskip 0.2cm

\centerline{ \em Department of Physics, Kyungpook National
University, Daegu, 702-701, Korea\footnote{Permanent address}}
\centerline{\em and} \centerline{\em Department of Physics, Syracuse
University, Syracuse, NY 13244-1130, USA.} \vskip 1.2cm

\begin{center}
{\bf Abstract}
\end{center}

Yang's theorem forbids the process $Z^0 \rightarrow 2\gamma$ in any
Poincar\'{e} invariant theory if photons are bosons and their
two-particle states transform under the Poincar\'{e} group in the
standard way (under the standard coproduct of the Poincar\'{e}
group). This is an important result as it does not depend on the
assumptions of quantum field theory. Recent work on noncommutative
geometry requires deforming the above coproduct by the Drinfel'd
twist. We prove that $Z^0 \rightarrow 2\gamma$ is forbidden for the
twisted coproduct as well. This result is also independent of the
assumptions of quantum field theory. As an illustration of the use
of our general formulae, we further show that $Z^0 \rightarrow \nu +
\nu$ is forbidden for the standard or twisted coproduct of the
Poincar\'{e} group if the neutrino is massless, even if lepton
number is violated. This is a special case of our general result
that a massive particle of spin $j$ cannot decay into two identical
massless particles of the same helicity if $j$ is odd, regardless of
the coproduct used.

 \vskip 1cm

\section{Introduction}

Many years ago, Yang \cite{Yang} proved the result that a massive
spin 1 particle cannot decay into two photons. The proof required
invariance under the Poincar\'{e} group ${\cal P}^{\uparrow}_{+}$
(without reflections), Bose statistics of photons and the assumption
that the two photon states transformed in the standard way under
${\cal P}^{\uparrow}_{+}$. (Many books \cite{gelfand, kahan,
naimark, wigner, balbook} treat the Poincar\'{e} group. See e.g.
Balachandran and Trahern \cite{trah} and references therein.)

Yang's proof does not use quantum field theory(QFT). It forbids the
decay $ Z^0 \rightarrow 2\gamma $. Limits on the branching ratio for
such processes thus give tests on the standard assumptions about
relativistic invariance and Bose symmetry which are insensitive to
models of QFT. This result of Yang is thus of basic significance.

Charge conjugation invariance does forbid the decay $ Z^0
\rightarrow 2\gamma $. But the standard model does not have this
invariance.

The structure of the Poincar\'{e} group ${\cal P}^{\uparrow}_{+}$
does not uniquely dictate the two-particle Poincar\'{e}
transformation law. If $x$ denotes spacetime coordinate and the
single particle wave functions $\psi$, $\chi$ transform according to
\beqa &\psi \rightarrow \Lambda \psi ,~~~\chi \rightarrow \Lambda
\chi ,\cr & &\cr & (\Lambda \psi )(x) := \psi (\Lambda^{-1}
x),~(\Lambda \chi )(x) := \chi (\Lambda^{-1} x) \eeqa under a
Lorentz transformation $\Lambda$, the two-particle wave function
$\psi \otimes \chi$ is customarily transformed according to \beqa
&\psi \otimes \chi \rightarrow (\Lambda
\otimes\Lambda)(\psi\otimes\chi) ,\cr & &\cr & (\Lambda
\otimes\Lambda)(\psi \otimes\chi)(x,y) = \psi (\Lambda^{-1}
x)~\chi(\Lambda^{-1} y ).\eeqa But this rule involves the choice of
a homomorphism $\Delta_0$ from the Lorentz group ${\cal
L}^{\uparrow}_{+}$ to ${\cal L}^{\uparrow}_{+} \times {\cal
L}^{\uparrow}_{+}$, namely, \be \bigtriangleup_0 (\Lambda) = \Lambda
\times \Lambda. \ee More generally, for the Poincar\'{e} group
${\cal P}^{\uparrow}_{+}$, we uncritically assume the homomorphism
\be \bigtriangleup_0 (g) = g \times g , ~~g \in {\cal
P}^{\uparrow}_{+}. \label{coproduct}\ee The choice of $\Delta_0$ is
not dictated by the Poincar\'{e} group and amounts to an additional
assumption.

The Poincar\'{e} group in fact admits more general coproducts and
hence more general transformation laws of multiparticle states.
These coproducts are parametrised by an antisymmetric matrix $\theta
= (\theta^{\mu \nu})$ with constant entries $\theta^{\mu \nu} = -
\theta^{\nu \mu}$ and are given by \beqa \Delta_{\theta} (g) &=&
F^{-1}_{\theta} (g \otimes g) F_{\theta}, \cr & &\cr F_{\theta} &=&
e^{-i P_{\mu} \otimes \theta^{\mu \nu} P_{\nu}}, \cr & &\cr P &=&
(P_{\mu}) : {\rm Four-momentum} . \eeqa $F_{\theta}$ is known as the
Drinfel'd twist \cite{Drin}. This twisted coproduct has become
central for the implementation of Poincar\'{e} invariance on the
Moyal plane \cite{chai, asch}.

The coproduct $\Delta_0$ defines the action of the Poincar\'{e}
group on multiparticle states.

It is clear from (\ref{coproduct}) that its action on two-particle
states commutes with the flip operator $\tau$: \be \tau ( \psi
\otimes \chi ) := \chi \otimes \psi . \ee Hence the subspaces with
elements \be P_{\pm} ~( \psi \otimes \chi),~ P_{\pm} = \frac{1}{2}
(1 \pm \tau )\ee are Poincar\'{e} invariant. Restriction to these
subspaces is thus compatible with Poincar\'{e} invariance. In this
way we are led to the concepts of bosons and fermions given by the
projectors $P_{\pm}$.

The transformation $\tau$ generalizes to $N$-particle sectors where
they generate the permutation group $S_N$. The projectors $P_{\pm}$
also generalize to $N$-particle sectors where they project to the
two one-dimensional representations of $S_N$.

But already at the two-particle level, the flip $\tau$ fails to
commute with $\Delta_{\theta} (g) $. Instead, we must replace $\tau$
by

\beqa \tau_{\theta} &=& F^{-1}_{\theta} \tau F_{\theta} ,
~~\tau^2_{\theta} = 1 \otimes 1 , \cr & &\cr \tau_0 &=& \tau
,\label{twistflip} \eeqa which {\it{does}} commute with
$\Delta_{\theta} (g)$ \cite{statistics, bal, lizzi, gauge}. The
twisted flip $\tau_{\theta}$ is associated with the new projectors
\beqa P^{\theta}_{\pm} &=& {1 \over 2} (1 \pm \tau_{\theta} ), \cr &
&\cr P^0_{\pm} &\equiv& P_{\pm} . \eeqa They define the twisted
bosonic and fermionic subspaces with elements $ P^{\theta}_{\pm}
~(\psi \otimes \chi)$.

The transformation $\tau_{\theta}$ as well generalizes to
$N$-particle sectors \cite{bal}.

In this paper, we first analyze the space of two-photon state
vectors for $\theta^{\mu \nu}=0$. It consists of vectors of the form
$P_{+} (\psi\otimes\chi)$. Using just group theory, we show that the
reduction of the representation of the Poincar\'{e} group ${\cal
P}^{\uparrow}_{+}$, acting by the coproduct $\Delta_0$ on this
space, does not contain its massive spin 1 representation. This
proves Yang's theorem.

Next, we repeat this analysis for the two-photon states given by the
projector $P^{\theta}_{+}$, the coproduct for ${\cal
P}^{\uparrow}_{+}$ being $\Delta_{\theta}$. We still find Yang's
result: This representation of the Poincar\'{e} group does not
contain the massive spin 1 representation. The process $Z^0
\rightarrow 2 \gamma$ is still forbidden. We show also that this
selection rule is a special case of a more general selection rule,
valid for any $\theta^{\mu \nu}$, forbidding the decay of a massive
particle of spin $j$ into two massless identical particles of the
same helicity if $j$ is odd.

Not all treatments of the standard model on the Moyal plane preserve
Poincar\'{e} invariance. The first treatment of $Z^0 \rightarrow 2
\gamma$ in a model violating Lorentz invariance is due to
\cite{wess}. More recent research on this subject can be found in
\cite{tram}. Also in the approach advocated by \cite{gauge}, based
on the twisted coproduct, for example, for reasons of locality, it
breaks down when a process involves both gauge and matter fields. In
this case, $Z^0 \rightarrow 2 \gamma$ need not be forbidden. Further
analysis of this approach is needed for a precise statement.

In the next two sections, we summarize the construction of the
unitary irreducible representations(UIRR's) of the universal
covering group ${\bar{\cal P}}^{\uparrow}_{+}$ of ${\cal
P}^{\uparrow}_{+}$ for massive and massless particles. (Not all zero
mass UIRR's are covered, only those of interest are described.)
Yang's theorem is then proved in section 4 and generalized to the
twisted coproduct case in section 5. Section 6 contains brief
concluding remarks.

\section{Irreducible Representations of $ {\cal P}^{\uparrow}_{+}$}

The Lie algebra of Poincar\'{e} group ${\cal P}^{\uparrow}_{+}$ is
spanned by the 10 generators $ J_{\mu \nu}$ and $P_\mu $ ($\mu , \nu
\in \{0,1,2,3\}$) which satisfy

\beqa [ J_{\alpha \beta},J_{\mu \nu} ]&=& i ( g_{\beta \mu}J_{\alpha
\nu} + g_{\alpha \nu}J_{\beta \mu} - g_{\alpha \mu}J_{\beta \nu} -
g_{\beta \nu}J_{\alpha \mu} ), \cr & &\cr [ J_{\alpha \beta},P_{\mu}
]&=& i ( g_{\beta \mu}P_{\alpha } + g_{\alpha \mu}P_{\beta }), \cr &
&\cr [ P_{\mu},P_{\nu} ]&=&0 . \eeqa  The Casimir operators of
${\cal P}^{\uparrow}_{+}$ are $ P^2 = P^{\mu} P_{\mu}$ and $ W^2 =
W^{\mu} W_{\mu}$ where $W_{\mu}=- {1 \over 2}\epsilon_{\mu \nu
\alpha \beta} J^{\nu \alpha}P^{\beta} $ is the Pauli-Lubanski
operator. These are represented by constants in irreducible
representations. We set $ P^2 = m^2 $ and consider only the cases
$m^2 \geq 0$ and $P_0 > 0$.

\subsection{Irreducible Representations for Massive Particles}

The construction of the UIRR's of ${\bar{\cal P}}^{\uparrow}_{+}$
are described in many books, for example in \cite{trah}. Here we
will briefly describe them.

For $m^2 > 0$ , the UIRR's of ${\bar{\cal P}}^{\uparrow}_{+}$ are
labeled by $m$ and $j$ with $ j=0, {1 \over 2}, 1, \cdots $. The
representation space of each UIRR is spanned by \{$ \mid p \ j
\lambda \rangle $\} where $p^{\mu} p_{\mu}=m^2$ and $\lambda = -j,
-j+1, \cdots, j-1, j$. Here, $p^{\mu}$ is a vector residing on the
three-dimensional hyperboloid $ \{ p \in R^4 \mid p^2 = m^2 , ~~p_0
> 0 \}$ and, consequently, the representation space is not
compact. This is natural because the group itself is not compact.
The basis states satisfy

\beqa P^{\mu} \mid p \ j \lambda \rangle &=& p^{\mu} \mid p \ j
\lambda \rangle , \cr & &\cr W^2 \mid p \ j \lambda \rangle &=& -m^2
j(j+1) \mid p \ j \lambda \rangle , \cr & &\cr \langle p' j'
\lambda' \mid pj\lambda \rangle &=&  2 p_0 \ \delta_{j' j} \
\delta_{\lambda' \lambda} \ \delta^3 (p'-p) .
\label{scalarmass}\eeqa

In order to understand the behavior of these states under the action
of an arbitrary Lorentz transformation, we have to be more precise
about the definition of the basis states.

For any given timelike 4-momentum $p^{\mu}$ with positive $p_0$,
there is a rest frame in which the momentum becomes
$\widehat{k}=(m,0,0,0)$. In this frame $\mid \widehat{k} \ j \lambda
\rangle $ is defined as a state satisfying

\beqa P^{\mu} \mid \widehat{k} \ j \lambda \rangle , &=&
{\widehat{k}}^{\mu} \mid \widehat{k} \ j \lambda \rangle , \cr &
&\cr L^2 \mid \widehat{k} \ j \lambda \rangle &=& j(j+1) \mid
\widehat{k}\  j \lambda \rangle \cr & &\cr L_3 \mid \widehat{k} \ j
\lambda \rangle &=& \lambda \mid \widehat{k} \ j \lambda \rangle .
\label{reststate}\eeqa Here, $L_i = {1\over 2} \epsilon_{ijk}J_{jk}$
and $L^2 = L^2_1 + L^2_2 + L^2_3$. In the rest frame,  $\mid
\widehat{k} \ j \lambda \rangle $ transforms as usual under a
spatial rotation $R$: \be U(R)\mid \widehat{k} \ j \lambda \rangle =
D^j_{\lambda' \lambda}(R) \mid \widehat{k} \ j \lambda' \rangle ,
\label{rotation}\ee $D^j (R)$ being spin $j$ rotation matrices. Also
$R \in SU(2)$ if $j \in \{1/2, 3/2,\cdots \}$.


 Going back from $\widehat{k}=(m,0,0,0)$ to the given
$p^{\mu}$ is achieved by a Lorentz transformation. However, there
are many Lorentz transformations which fulfill this job. The
ambiguity comes from the existence of a non-trivial stability group
of $\widehat{k}$, which, in this case, is the rotation subgroup. We
fix the ambiguity by choosing the Lorentz transformation $L(p)$
which transforms $\widehat{k}$ to $p$, i.e. $p=L(p)\widehat {k}$, as
follows:

\be L(p)= e^{-i\alpha J_{12}}e^{-i\beta J_{31}}e^{i\alpha
J_{12}}e^{-i\delta J_{03}}. \label{boost}\ee The values of
$\alpha,\beta$ are fixed by the spatial part of $p^\mu$ and that of
$\delta$ is fixed by the time component of $p^\mu$. With this
$L(p)$, we define our general basis state $ \mid p \ j \lambda
\rangle $ by

\be  \mid p \ j \lambda \rangle = U(L(p)) \mid \widehat{k} \ j
\lambda \rangle . \label{genstate}\ee

In order to see how $\mid p \ j \lambda \rangle$ transforms under an
arbitrary Lorentz transformation $\Lambda$, we consider

\beqa U(\Lambda)\mid p  \  j \lambda \rangle &=& U(L(\Lambda p))
U(L^{-1}(\Lambda p)) U(\Lambda) U(L(p))\mid \widehat{k} \ j \lambda
\rangle \cr & &\cr &=& U(L(\Lambda p)) U(L^{-1}(\Lambda p) \Lambda
L(p))\mid \widehat{k} \ j \lambda \rangle . \eeqa Here, $L(\Lambda
p)$ is the Lorentz transformation of the form given in
(\ref{boost}), which maps $\widehat{k}$ to $\Lambda p$. Notice that
$L^{-1}(\Lambda p) \Lambda L(p)$ leaves $\widehat {k} $ invariant.
Therefore, it must be a pure spatial rotation. We denote it by
$R(\Lambda,p)$. Using (\ref{rotation}), we get

\be U(\Lambda)\mid p \ j \lambda \rangle = D^j_{\lambda'
\lambda}(R(\Lambda,p)) \mid \Lambda p \ j \lambda' \rangle . \ee We
see that the first two equations in (\ref{scalarmass}) can be
derived using (\ref{reststate}) and (\ref{genstate}).

This representation of the Poincar\'{e} group is unitary for the
scalar product given by (\ref{scalarmass}).

We denote the vector space spanned by $\{ \mid p \ j \lambda \rangle
\}$ as $V(\lambda )$.

\subsection{Irreducible Representations for Massless Particles}

Now we consider the case $m=0$. In this case, the UIRR's of ${\bar
{\cal P}}^{\uparrow}_{+}$ are characterized by a continuous
parameter $\rho$ with $ 0\leq \rho < \infty $ and the sign of
energy(sign $p_0$).

For a given $\rho$ with $ \rho > 0 $ and a given sign $p_0$, there
are two irreducible representations. The representation space is
spanned by \{$ \mid p \  \lambda \ \rho \ ({\rm {sign}}\ p_0 )
\rangle $\} with $p^{\mu} p_{\mu}=0$. For the first irreducible
representation, $\lambda = \cdots, -1, 0, 1, \cdots $, while for the
second irreducible representation, $\lambda = \cdots, -{1 \over 2} ,
{1 \over 2}, {3 \over 2}, \cdots $. Under $2 \pi $ rotation, the
first set of states are invariant while the second states change
sign. The basis states satisfy

\beqa P^{\mu} \mid p \  \lambda \ \rho \ ({\rm {sign}}\ p_0 )
\rangle , &=& p^{\mu} \mid p \  \lambda \ \rho \ ({\rm {sign}}\ p_0
) \rangle \cr & &\cr W^2 \mid p \  \lambda \ \rho \ ({\rm {sign}}\
p_0 ) \rangle &=& -{\rho}^2 \mid p \  \lambda \ \rho \ ({\rm
{sign}}\ p_0 ) \rangle . \label{masslessstate}\eeqa We skip the
analysis of the behavior of these states under an arbitrary Lorentz
transformation.

For $\rho = 0$, there are an infinite number of inequivalent UIRR's.
They are labelled by helicity $\lambda$ with $\lambda \in \{\cdots,
-1, -{1 \over 2} ,0, {1 \over 2}, 1, \cdots \}$ and by sign $p_0$.
We fix sign $p_0$ to be positive as that is the case of interest.
Each representation space is then spanned by \{$ \mid p \ \lambda
\rangle \mid p^2 = 0, p_0
> 0 $\} for a fixed $\lambda$. Note that distinct $\lambda$
define inequivalent irreducible representations of ${\bar {\cal
P}}^{\uparrow}_{+}$.

Photons are described by the UIRR's with $\rho=0$ and $\lambda=\pm
1$. Integral values of $\lambda$ give UIRR's of ${\cal
P}^{\uparrow}_{+}$.


Let us discuss the behavior of $ \mid p \ \lambda \rangle $ under
the action of an arbitrary Lorentz transformation. For any
light-like four-momentum $p^{\mu}$ with positive $p_0$, there is a
frame in which the momentum becomes
$\widehat{k}=(\omega,0,0,\omega)$. The stability group of $\widehat
{k}$ is the group generated by \{ $\Pi_1 , \ \Pi_2 , \ L_3 $\} where
$\Pi_1 = J_{10} - J_{13}$ and $\Pi_2 = J_{20} - J_{23}$. Their
commutation relations are

\beqa [ L_3 ,\ \Pi_1 ] &=& i\ \Pi_2 , \cr & &\cr [ L_3 , \ \Pi_2
]&=& -i \ \Pi_1 , \cr & &\cr [\Pi_1 , \ \Pi_2 ] &=& 0 . \eeqa This
group is isomorphic to the Euclidean group in two dimensions. In the
frame where the four-momentum is $ \widehat {k}^\mu$, $\mid
\widehat{k} \ \lambda \rangle $ is defined as a state satisfying

\beqa P^{\mu} \mid \widehat{k} \  \lambda \rangle &=&
{\widehat{k}}^{\mu} \mid \widehat{k} \  \lambda \rangle , \cr & &\cr
L_3 \mid \widehat{k} \  \lambda \rangle , &=& \lambda \mid
\widehat{k} \  \lambda \rangle \cr & &\cr \Pi_i \mid \widehat{k} \
\lambda \rangle &=& 0.  \label{stabilityaction}\eeqa

As in the massive case, we introduce a Lorenz transformation $L(p)$
of the form (\ref{boost}), which maps $\widehat{k}$ to a given
light-like 4 momentum $p^{\mu}$. With this $L(p)$, $ \mid p \
\lambda \rangle $ is defined as

\be  \mid p \  \lambda \rangle = U(L(p)) \mid \widehat{k} \ \lambda
\rangle . \ee

Under an arbitrary Lorentz transformation $\Lambda$, we have

\beqa U(\Lambda)\mid p  \   \lambda \rangle &=& U(L(\Lambda p))
U(L^{-1}(\Lambda p) \Lambda L(p))\mid \widehat{k} \  \lambda \rangle
, \eeqa where $L^{-1}(\Lambda p) \Lambda L(p)$ is an element of the
stability group of $\widehat{k}=(\omega,0,0,\omega)$. The action of
the stability group on $\mid \widehat{k} \  \lambda \rangle$ is
given in (\ref{stabilityaction}). Therefore, the above equation is
equal to $\mid \Lambda p  \ \lambda \rangle$ times a phase factor.

We normalize the states by

\be \langle p' \ \lambda' \mid p \  \lambda \rangle = 2 p_0 \
 \delta_{\lambda' \lambda} \
\delta^3 (p'-p) . \label{scalarless}\ee

Using (\ref{masslessstate}) and (\ref{scalarless}), we can show that
the above representations for $m=0$ are unitary.

\section{Reduction of the Direct Product of Two Massless States: No Twist}

The direct product of two UIRR's of the Poincar\'{e} group can be
reduced into a direct sum of UIRR's. We consider the product of two
massless representations. Here, we exclude $\rho \neq 0$ and ${\rm
{sign}} \ p_0 < 0 $ massless representations. The product states are
then massive except when two massless states have parallel momenta.
In this exceptional case, the product representation is also
irreducible:

\be \mid p_1 \  \lambda_1 \rangle \mid p_2 \  \lambda_2 \rangle \sim
\mid p_1 + p_2 \  \lambda_1 + \lambda_2 \rangle . \ee Note that this
relation is defined upto a normalization factor. We do not consider
this case further. It does not affect the process $Z^0 \rightarrow
2\gamma$.

We consider a two massless-particle state with fixed helicities
$\lambda_i$ $(i=1,2)$. A general state can be expressed as a linear
sum of the basis states $\{\mid p_1 \ \lambda_1 \rangle \mid p_2 \
\lambda_2 \rangle \ \}$. The representation space
$V(\lambda_1)\otimes V(\lambda_2 )$ spanned by the basis is
irreducible with respect to the direct product of the two
Poincar\'{e} groups. However, under the diagonal subgroup, this
space is reducible.

The reduction of the direct product of two massless representations
can be summarized by the following formula:

\be \mid \lambda_1 \lambda_2 \ \widehat{p} \ j \mu \rangle =
\int_{SU(2)} d \mu (R) {D^{j *}_{\mu \ \lambda_1 - \lambda_2}} (R)
\Delta_0 (R) \mid q_1 \ \lambda_1 \rangle \mid q_2 \ \lambda_2
\rangle. \label{massivered}\ee Here, $d \mu (R)$ is the invariant
Haar measure on the $SU(2)$ group manifold. It is normalized by
$\int_{SU(2)}d \mu (R) = 1$. The momenta of the two particles are
fixed by $q_1 = (q, 0,0, q) $ and $q_2 = (q,0,0,-q)$ with positive
$q$. Therefore, the state is described in the center of momentum
frame and $\widehat{p}=(M,0,0,0)$ with $M=2q$ as the mass of the two
particle system.

We can understand this crucial formula as follows. We have to verify
that the left-hand side transforms under $SU(2)$ like a vector with
angular momentum $j$ and its third component $\mu$. Now under $S \in
SU(2)$, $\mid \lambda_1 \lambda_2 \ \widehat{p} \ j \mu \rangle $
transforms to $ \int_{SU(2)} d \mu (R)~~ {D^{j
*}_{\mu \ \lambda_1 - \lambda_2}} (R) ~~\Delta_0 (S)\\
\Delta_0 (R) \mid q_1 \ \lambda_1 \rangle \mid q_2 \ \lambda_2
\rangle $. Using $\Delta_0 (S) \Delta_0 (R) = \Delta_0 (SR)$ and the
invariance of the measure, the transformed state can be shown to be
$D^j_{\alpha \mu} (S) \mid \lambda_1 \lambda_2 \ \widehat{p} \ j
\alpha \rangle $, which verifies the validity of (\ref{massivered}).

The state in an arbitrary frame can be obtained by the corresponding
Lorentz transformation as in the single particle case:

\be \mid \lambda_1 \lambda_2 \ p \ j \mu \rangle \;=\; \Delta_0
(L(p)) \mid \lambda_1 \lambda_2 \ \widehat{p} \ j \mu \rangle.
\label{generalmom}\ee

It can be shown that the states $\mid \lambda_1 \lambda_2 \
\widehat{p} \ j \mu \rangle$ with $\mu = -j, -j+1, \cdots , j-1, j $
and their Lorentz transforms form a basis for a UIRR labelled by
$\{\lambda_1 , \lambda_2 , M, j \}$. We denote the space as
$\tilde{V}(\lambda_1 , \lambda_2 , M, j )$. It can also be shown
that any state in $V(\lambda_1)\otimes V(\lambda_2 )$ can be
expressed as a superposition of $\mid \lambda_1 \lambda_2 \ p \ j
\mu \rangle$ with different $\{M, j \}$. It shows that

\be V(\lambda_1)\otimes V(\lambda_2 ) = \bigoplus_{M, j}
\tilde{V}(\lambda_1 , \lambda_2 , M, j ). \ee On the right hand side
of this expression, the value of $M$ runs over all positive values
and the value of $j$ is lower-bounded by $\mid \lambda_1 -\lambda_2
\mid$.

Note that we have considered only the cases $M>0$ in the above
discussion.

In order to obtain Clebsch-Gordan coefficients, we write, for $R \in
SU(2)$,

\beqa R &=& e^{-i \alpha J_{12}} \  e^{-i \beta J_{31}} \ e^{-i
\gamma J_{12}},\cr & &\cr d \mu (R) &=&  {1 \over {16 \pi^2 }}~d
\alpha ~d cos \beta ~d \gamma ,~ \alpha \in [0, 2\pi ], ~\beta \in
[0, \pi ], ~\gamma \in [0, 4 \pi ]. \label{coord} \eeqa Then,
(\ref{massivered}) becomes

\be \mid \lambda_1 \lambda_2 \ \widehat{p} \ j \mu \rangle = \frac
1{4\pi} \int_0^{2\pi} d \alpha \int_{-1}^1 \;d\cos \beta \; {d^j
}_{\mu, \lambda_1 - \lambda_2}(\beta(\vec{p_1})) \;
e^{i(\mu-\lambda_1 -\lambda_2 )\alpha(\vec{p_1})} \mid p_1 \
\lambda_1 \rangle \mid p_2 \ \lambda_2 \rangle_{\rm{CM}}.
\label{massiveredf}\ee Here, $p_1 = (p_{10}, \vec{p_1})$ and ${d^j
}_{\mu, \lambda_1 - \lambda_2}(\beta)= D^j_{\mu , \lambda_1 -
\lambda_2 } (e^{-i \beta J_{31}})$. Coordinates $(\alpha(\vec{p_1})
, \beta(\vec{p_1}))$ are the azimuthal and polar angles of $\vec
{p_1}$. The subscript `$\rm{CM}$' denotes the `center-of-momentum'
frame where $p_2 = (p_{20}, \vec{p_2})$ with
$\vec{p_1}+\vec{p_2}=0$. Therefore, the corresponding angles of
$\vec{p_2}$ are $(\alpha + \pi , \pi - \beta)$.

The conventions (\ref{boost}) and (\ref{genstate}) for defining the
basis state have to be carefully followed to obtain
(\ref{massiveredf}). We illustrate how the calculation is done for
the factors involving $\lambda_2$ in (\ref{massiveredf}). First note
that the $\gamma$ dependent terms in (\ref{massivered}) cancel out.
So we focus on the relevant term coming from $\Delta_0 (R)$ and
$\mid q_2\ \lambda_2 \rangle$. It is

\beqa e^{-i \alpha({\vec p_1}) J_{12}} e^{-i \beta({\vec p_1})
J_{31}} \mid q_2 \ \lambda_2 \rangle &=& e^{-i \alpha({\vec p_1})
J_{12}} e^{-i \beta({\vec p_1}) J_{31}} e^{-i \pi J_{31}}\mid q_1 \
\lambda_2 \rangle \cr & &\cr &=& e^{-i (\alpha({\vec p_1}) + \pi )
J_{12}} e^{i \beta({\vec p_1}) J_{31}} e^{i \pi J_{12}} e^{-i \pi
J_{31}}\mid q_1 \ \lambda_2 \rangle \cr & &\cr &=& e^{-i
(\alpha({\vec p_1}) + \pi ) J_{12}} e^{-i(\pi- \beta({\vec p_1}))
J_{31}} e^{i (\alpha({\vec p_1}) + \pi) J_{12}} e^{-i (\alpha({\vec
p_1}) + 2\pi) J_{12}}\mid q_1 \ \lambda_2 \rangle \cr & &\cr &=&
e^{-i (\alpha({\vec p_1}) +2 \pi ) \lambda_2} \mid p_2 \ \lambda_2
\rangle = (-1)^{2\lambda_2} e^{-i\alpha({\vec p_1})\lambda_2 }\mid
p_2 \ \lambda_2 \rangle. \eeqa The factor $(-1)^{2\lambda_2}$ is an
overall factor and will be absorbed into a new definition of the
state $\mid \lambda_1 \lambda_2 \ \widehat{p} \ j \mu \rangle$. The
$\lambda_2$-dependence of the second index in ${d^j }_{\mu,
\lambda_1 - \lambda_2}(\beta(\vec{p_1}))$ comes directly from ${D^{j
*}_{\mu \ \lambda_1 - \lambda_2}} (R)$ in (\ref{massivered}). We
thus account for the $\lambda_2$-terms in (\ref{massiveredf}).

Inverting (\ref{massivered}) we get

\be \Delta_0 (R) \mid q_1 \  \lambda_1 \rangle \mid q_2 \ \lambda_2
\rangle \;=\; \sum_{j,\mu} (2j+1) D^j_{\mu , \lambda_1 -
\lambda_2}(R) \mid \lambda_1 \lambda_2 \widehat{p} \ j \mu \rangle
.\ee From this and using (\ref{coord}) we have

\be \mid p_1 \  \lambda_1 \rangle \mid p_2 \  \lambda_2
\rangle_{\rm{CM}}\;=\;\sum_{j,\mu} (2j+1) e^{-i(\mu-\lambda_1
-\lambda_2 )\alpha({\vec {p_1} })} {d^j }_{\mu, \lambda_1 -
\lambda_2}(\beta(\vec {p_1})) \mid \lambda_1 \lambda_2 \ \widehat{p}
\ j \mu \rangle . \label{massiveredinvf}\ee The Clebsch-Gordan
coefficients in the center-of-momentum frame are determined by
(\ref{massiveredf}) and (\ref{massiveredinvf}). Relations in the
general frame can be obtained by Lorentz transforming these two
equations. We thus get

\beqa \langle k_1 \  \lambda_1  \mid \langle k_2 \  \lambda_2 \mid
\lambda_1 \lambda_2 \widehat{p} \ j \mu \rangle \;=\; & \frac
{1}{\pi } d^j_{\mu , \lambda_1 - \lambda_2 } [\beta(\vec{k_1})]
e^{i(\mu - \lambda_1 - \lambda_2) \alpha(\vec{k_1})} \cr & &\cr &
 \delta(\mid \vec{k_1}\mid -
q ) \delta^3 ( \vec{k_1}+ \vec{k_2}) ,\label{partinner}\eeqa and

\be \langle \lambda_1 \lambda_2 p' \ j' \mu' \mid \lambda_1
\lambda_2 p \ j \mu \rangle \;=\; \frac {2}{\pi (2j+1)} \delta_{j' j
} \delta_{\mu' \mu} \delta^4 ( p' - p). \label{innerproduct}\ee We
can get (\ref{innerproduct}) quickly as follows. All but the overall
normalization factor $ 2/\pi(2j+1)$ in (\ref{innerproduct}) is fixed
by general considerations. To get the overall factor, we put
$p=\widehat{p}$ and use (\ref{partinner}). Then (\ref{partinner})
vanishes unless $\langle k_1 \  \lambda_1  \mid \langle k_2 \
\lambda_2 \mid$ is ${}_{CM}\langle p_1 \  \lambda_1  \mid \langle
p_2 \  \lambda_2 \mid$. Substituting for the former in
(\ref{partinner}) by the latter from (\ref{massiveredinvf}), we get
the factor $ 2/\pi(2j+1)$ in (\ref{innerproduct}). The factor 2
comes because the total center-of-momentum energy is twice the
energy of either particle and $\delta(x)=2\delta(2x)$.

\section{The Case of Two Identical Particles}

When we consider two identical particles, the product state must be
either symmetrized or anti-symmetrized depending on the spin of the
particle. The reduction formula should be modified accordingly. For
the case of massless particles, we get

\be \mid \lambda_1 \lambda_2 \ \widehat{p} \ j \mu \rangle_{S,A} =
\int_{SO(3)} d \mu (R) {D^* (R)}^j_{\mu \ \lambda_1 - \lambda_2}
\Delta_0 (R) \frac{1 \pm \tau}{2} \mid q_1 \ \lambda_1 \rangle \mid
q_2 \ \lambda_2 \rangle. \label{symmassivered}\ee Here, $\tau$ is
the flip operator, \be \tau \mid q_1 \lambda_1 \rangle \mid q_2
\lambda_2 \rangle = \mid q_2 \lambda_2 \rangle \mid q_1 \lambda_1
\rangle , \ee and $S(A)$ denotes the symmetric (anti-symmetric)
state. We take $+$ if the particles are tensorial (their helicities
are integral) and we take $-$ if they are spinorial (their
helicities are $\pm 1/2 , \pm 3/2 , \cdots $). Note here that the
two helicities $\lambda_1$ and $\lambda_2$ may be different.
Massless particle states with different helicities never mix under
the Poincar\'{e} group ${\cal P}^{\uparrow}_{+}$. However, the
disconnected component of the Poincar\'{e} group will mix different
helicity states. For example, under parity, helicity changes sign so
that the helicity of the photon can be $ \pm 1$.

The coproduct $\Delta_0 (R)$ and $\tau$ commute and we can write

\be \mid \lambda_1 \lambda_2 \ \widehat{p} \ j \mu
\rangle_{S,A}\;=\; \frac{1 \pm \tau}{2} \mid \lambda_1 \lambda_2 \
\widehat{p} \ j \mu \rangle . \ee

The action of $\tau$ on $\mid \lambda_1 \lambda_2 \ \widehat{p} \ j
\mu \rangle$ changes the order of the two one-particle states and we
get

\be \tau \mid \lambda_1 \lambda_2 \ \widehat{p} \ j \mu \rangle =
\frac 1{4\pi} \int_0^{2\pi} d \alpha \int_{-1}^1 \;d\cos \beta \;
{d^j }_{\mu, \lambda_1 - \lambda_2}(\beta) \; e^{i(\mu - \lambda_1
-\lambda_2 )\alpha} \mid p_2 \  \lambda_2 \rangle \mid p_1 \
\lambda_1 \rangle_{\rm{CM}}. \ee Here, the momenta of two particles
are given by $p_1 = (q,\vec{p_1}) $ and $p_2 = (q,- \vec{p_1})$ with
the direction of $\vec{p_1}$ denoted by $(\alpha, \beta)$.
Identifying $\mid p_1 \  \lambda_1 \rangle$ by $\mid
\overrightarrow{(\alpha,\beta)} \  \lambda_1 \rangle$, we have

\be \mid p_2 \ \lambda_2 \rangle \mid p_1 \  \lambda_1
\rangle_{\rm{CM}}\;=\; \mid -\overrightarrow{(\alpha,\beta)}\
\lambda_2 \rangle \mid \overrightarrow{(\alpha,\beta)} \ \lambda_1
\rangle . \ee Using
$-\overrightarrow{(\alpha,\beta)}=\overrightarrow{(\alpha +\pi,\pi -
\beta)}$, the above state can be written as

\be \mid p_2 \  \lambda_2 \rangle \mid p_1 \  \lambda_1
\rangle_{\rm{CM}}\;=\; \mid \overrightarrow{(\alpha +\pi,\pi
-\beta)}\  \lambda_2 \rangle \mid  - \overrightarrow{(\alpha + \pi,
\pi - \beta)} \  \lambda_1 \rangle . \ee We now change the
integration variables from $\alpha$ and $\beta$ to $\tilde{\alpha}
=\alpha + \pi$ and $\tilde{\beta}=\pi - \beta$ and get

\beqa \tau \mid \lambda_1 \lambda_2 \ \widehat{p} \ j \mu \rangle\;
=\; & (-1)^{j+\lambda_1 +\lambda_2}\frac 1{4\pi} \int_0^{2\pi} d
\tilde{\alpha} \int_{-1}^1 \;d\cos \tilde{\beta} \; \cr & &\cr &{d^j
}_{\mu, \lambda_2 - \lambda_1}(\tilde{\beta}) \; e^{i(\mu-\lambda_1
-\lambda_2 )\tilde{\alpha}} \mid
\overrightarrow{(\tilde{\alpha},\tilde{\beta)}} \  \lambda_2 \rangle
\mid -\overrightarrow{(\tilde{\alpha},\tilde{\beta)}} \ \lambda_1
\rangle. \eeqa Here, we have used the identity:

\be \;d^j_{\mu \nu} (\pi-\beta)\;=\; (-1)^{(j+\mu)} d^j_{\mu (-\nu)}
(\beta). \ee  This identity is well-known in angular momentum theory
\cite{angular}. Comparing this with (\ref{massiveredf}), we have,
\be \tau \mid \lambda_1 \lambda_2 \ \widehat{p} \ j \mu \rangle\;
=\;(-1)^{j+\lambda_1 +\lambda_2}\mid \lambda_2 \lambda_1 \
\widehat{p} \ j \mu \rangle , \ee and therefore,

\be \mid \lambda_1 \lambda_2 \ \widehat{p} \ j \mu
\rangle_{S,A}\;=\; \frac{1}{2} \left( \mid \lambda_1 \lambda_2 \
\widehat{p} \ j \mu \rangle \pm (-1)^{(j+\lambda_1 +\lambda_2 )}\mid
\lambda_2 \lambda_1 \ \widehat{p} \ j \mu \rangle \right) .
\label{symred}\ee

This equation determines the selection rules. For example, Yang's
argument about the forbidden decay of $Z^0 \rightarrow 2 \gamma$ can
be easily explained using this equation as follows.

The particle $Z^0$ has spin $j=1$. Therefore, the two photons after
the $Z^0$ decay at rest cannot have opposite helicities by angular
momentum conservation. For if the two photons have opposite
helicities, then $\mid \lambda_1 - \lambda_2 \mid = 2$ and the
minimum value for $j$ is 2. This is bigger than the spin of $Z^0$
which is 1.

Now we assume that the two photons after decay have the same
helicity, that is, $\lambda_1 = \lambda_2 =\lambda$. In this case,
(\ref{symred}) becomes

\be \mid \lambda \lambda \ \widehat{p} \ j \mu \rangle_S\;=\;
\frac1{2} \left(1 + (-1)^{(j+\lambda +\lambda )}\right)  \mid
\lambda \lambda \ \widehat{p} \ j \mu \rangle  .
\label{photonred}\ee We choose $+$ because photon is a boson. Now
substituting $j=1$ and $\lambda = \pm 1$, we find that the right
hand side vanishes. This means that two photon states cannot have
any $j=1$ component. Consequently, the decay $Z^0 \rightarrow 2
\gamma$ is forbidden.

So far, we have considered the standard coproduct of the
Poincar\'{e} group acting on the tensor product states. In the next
section, we introduce a new coproduct and investigate how to reduce
the direct product of two irreducible representations with this new
coproduct.

\section{Twisted Coproduct}

We now replace the coproduct $\Delta_0 (R)$ by the twisted coproduct
$\Delta_{\theta} (R)$ to define a new action of Poincar\'{e}
transformation on the direct product states as was discussed in the
introduction. The direct product of two irreducible representations
of the Poincar\'{e} group is also reducible under the action of this
twisted coproduct. The way to reduce the direct product space is the
same as in the untwisted coproduct case except that the untwisted
coproduct $\Delta_0 (R)$ should be replaced by the twisted coproduct
$\Delta_{\theta} (R)$. For the case of two massless particle
systems, we have

\be \mid \lambda_1 \lambda_2 \ \widehat{p} \ j \mu \rangle_{\theta}
= \int_{SU(2)} d \mu (R) D^{j*}_{\mu \ \lambda_1 - \lambda_2} (R)
\Delta_{\theta} (R) \mid q_1 \  \lambda_1 \rangle \mid q_2 \
\lambda_2 \rangle. \label{twistmassivered}\ee It can be shown that
the subspace generated by the above states forms an irreducible
subspace under the twisted coproduct action of the Poincar\'{e}
group. That is, the state $\mid \lambda_1 \lambda_2 \ \widehat{p} \
j \mu \rangle_{\theta}$ transforms under the action of the twisted
coproduct of the Poincar\'{e} group as if it is a single particle
state with mass $2q$ and spin $j$ just like the way that $\mid
\lambda_1 \lambda_2 \ \widehat{p} \ j \mu \rangle$ transforms under
the action of the untwisted coproduct. It can also be shown that the
collection of $\{\mid \lambda_1 \lambda_2 \ \widehat{p} \ j \mu
\rangle_{\theta}\}$ and their Lorentz transformations with different
$\lambda_1, \lambda_2, j$ form a complete set for the direct product
space. Note here that the two particle state on the right hand side
of (\ref{twistmassivered}) is taken to be the ordinary tensor
product state. If we use the star(or twisted) tensor product state
instead defined by \cite{lizzi, finland}

\be \mid \Psi \rangle {\otimes}_{\theta} \mid \Phi \rangle =
F^{-1}_{\theta} \mid \Psi \rangle {\otimes} \mid \Phi \rangle, \ee
there will be an extra overall phase factor on the right hand side
of (\ref{twistmassivered}), which is quite irrelevant in the
following arguments.

The action of the twisted coproduct on the tensor product state is

\be \Delta_{\theta} (g) \mid q_1 \ \lambda_1 \rangle \mid q_2 \
\lambda_2 \rangle\;=\; e^{-\frac{i}{2}q_1 \wedge q_2}
F^{-1}_{\theta} \Delta_0 (g)\mid q_1 \ \lambda_1 \rangle \mid q_2 \
\lambda_2 \rangle ,\ee and therefore

\be \mid \lambda_1 \lambda_2 \ \widehat{p} \ j \mu \rangle_{\theta}
\; =\; e^{-\frac{i}{2}q_1 \wedge q_2}F^{-1}_{\theta} \mid \lambda_1
\lambda_2 \ \widehat{p} \ j \mu \rangle. \ee Here, $p \wedge q
=p_\mu \theta^{\mu \nu} q_\nu $. Substituting (\ref{massiveredf}) in
this equation, we get

\beqa \mid \lambda_1 \lambda_2 \ \widehat{p} \ j \lambda
\rangle_{\theta}\; =\; & \frac {1}{4\pi} \int_0^{2\pi} d \alpha
\int_{-1}^1 \;d\cos \beta \; {d^j }_{\mu, \lambda_1 -
\lambda_2}(\beta) \; e^{i(\mu-\lambda_1 -\lambda_2 )\alpha} \cr &
&\cr & e^{\frac{i}{2}(p_1 \wedge p_2 - q_1 \wedge q_2)} \mid p_1 \
 \lambda_1 \rangle \mid p_2 \  \lambda_2 \rangle_{\rm{CM}}.
\eeqa If $\theta^{0i}=0$, then since $\vec{p_1 },~\vec{p_2}
(\vec{q_1 },~\vec{q_2 })$ are antiparallel in the center-of-momentum
frame, $p_1 \wedge p_2 = q_1 \wedge q_2 =0$ and $\mid \lambda_1
\lambda_2 \ \widehat{p} \ j \mu \rangle_{\theta}$ and $\mid
\lambda_1 \lambda_2 \ \widehat{p} \ j \mu \rangle$ are identical.
However, using (\ref{generalmom}) the twisted state in an arbitrary
frame is seen to be

\be \mid \lambda_1 \lambda_2 \ p \ j \mu \rangle_{\theta} \;=\;
e^{-\frac{i}{2}q_1 \wedge q_2} F^{-1}_{\theta}\mid \lambda_1
\lambda_2 \ p \ j \mu \rangle, \label{twistedstate}\ee so that $\mid
\lambda_1 \lambda_2  p \ j \mu \rangle_{\theta}$ and $\mid \lambda_1
\lambda_2 \ p \ j \mu \rangle$ will in general be different if
$\theta^{ij}\neq 0$ even if $\theta^{0i}=0$.

The Clebsch-Gordan coefficients are modified:

\be \mid p_1 \ \lambda_1 \rangle \mid p_2 \ \lambda_2
\rangle_{\rm{CM}}\;=\; e^{\frac{i}{2}(q_1 \wedge q_2 - p_1 \wedge
p_2 )}\sum_{j,\mu} (2j+1) e^{-i(\mu-\lambda_1 -\lambda_2 )\alpha}
{d^j }_{\mu, \lambda_1 - \lambda_2}(\beta) \mid \lambda_1 \lambda_2
\ \widehat{p} \ j \mu \rangle_{\theta} ,\ee

\be \langle k_1 \  \mu_1  \mid \langle k_2 \  \mu_2 \mid \lambda_1
\lambda_2 \widehat{p} \ j \mu \rangle_{\theta} \;=\;
e^{\frac{i}{2}(k_1 \wedge k_2 - q_1 \wedge q_2 )} \langle k_1 \
\mu_1  \mid \langle k_2 \  \mu_2 \mid \lambda_1 \lambda_2
\widehat{p} \ j \mu \rangle ,\ee

\be{}_{\theta} \langle \lambda'_1 \lambda'_2 p' \ j' \mu' \mid
\lambda_1 \lambda_2 p \ j \mu \rangle_{\theta} \;=\; \langle
\lambda'_1 \lambda'_2 p' \ j' \mu' \mid \lambda_1 \lambda_2 p \ j
\mu \rangle .\ee

Finally, we discuss the tensor product of two identical particle
states. With the twisted coproduct, symmetrization or
antisymmetrization should be done not with $\tau$ but with
$\tau_{\theta}$ defined in the introduction. With this twisted flip
operator, we get

\be \mid \lambda_1 \lambda_2 \ \widehat{p} \ j \mu
\rangle^{S,A}_{\theta} \;=\; \frac{1 \pm \tau_{\theta}}{2} \mid
\lambda_1 \lambda_2 \ \widehat{p} \ j \mu \rangle_{\theta} . \ee
Substituting (\ref{twistmassivered}) into above equation, we obtain
\be \mid \lambda_1 \lambda_2 \ \widehat{p} \ j \mu
\rangle^{S,A}_{\theta} = \int_{SU(2)} d \mu (R) D^{j*}_{\mu \
\lambda_1 - \lambda_2} (R) \frac{1 \pm \tau_{\theta}}{2}
\Delta_{\theta} (R) \mid q_1 \ \lambda_1 \rangle \mid q_2 \
\lambda_2 \rangle. \label{twistmassivered2}\ee Using the relations
$1\pm \tau_{\theta} = F^{-1}_{\theta} (1\pm \tau ) F_{\theta}$ and
$\Delta_{\theta}(R)=F^{-1}_{\theta} \Delta_0 (R) F_{\theta}$, we get
\be \mid \lambda_1 \lambda_2 \ \widehat{p} \ j \mu
\rangle^{S,A}_{\theta} =e^{- {\frac{i}{2}}q_1 \wedge q_2}
F^{-1}_{\theta} \int_{SU(2)} d \mu (R) D^{j*}_{\mu \ \lambda_1 -
\lambda_2} (R) \frac{1 \pm \tau}{2} \Delta (R) \mid q_1 \ \lambda_1
\rangle \mid q_2 \ \lambda_2 \rangle. \label{twistmassivered3}\ee
Comparing this result with (\ref{symmassivered}), we obtain \be \mid
\lambda_1 \lambda_2 \ \widehat{p} \ j \mu \rangle^{S,A}_{\theta}
=e^{- {\frac{i}{2}}q_1 \wedge q_2} F^{-1}_{\theta} \mid \lambda_1
\lambda_2 \ \widehat{p} \ j \mu \rangle^{S,A}\ee and \be \mid
\lambda_1 \lambda_2 \ \widehat{p} \ j \mu
\rangle^{S,A}_{\theta}\;=\; \frac{1}{2} \left( \mid \lambda_1
\lambda_2 \ \widehat{p} \ j \mu \rangle_{\theta} \pm
(-1)^{(j+\lambda_1 +\lambda_2 )}\mid \lambda_2 \lambda_1 \
\widehat{p} \ j \mu \rangle_{\theta} \right). \label{aaa}\ee Here we
used (\ref{twistedstate}). In case $\lambda_1 =\lambda_2=\lambda$,
we thus have

\be \mid \lambda \lambda \ \widehat{p} \ j \mu
\rangle^{S,A}_{\theta} \;=\; \frac{1}{2} \left( 1 \pm (-1)^{(j+
2\lambda )}\right) \mid \lambda \lambda \ \widehat{p} \ J \mu
\rangle_{\theta} , \label{selectionrule}\ee and consequently, the
selection rules are not altered by twisting the coproduct. The decay
$Z_0 \rightarrow 2 \gamma$ is forbidden even with the twisted
coproduct. Note that this result is somehow expected because the
twist operator carries only momentum (and no spin) degrees of
freedom. The relative phase in (\ref{aaa}), and consequentely in
(\ref{selectionrule}), is not altered by the introduction of the
twist.

Equation (\ref{selectionrule}) shows that a massive particle of spin
$j$ cannot decay into a pair of identical massless particles of
helicity $\lambda$ if $1+(-1)^{2\lambda} (-1)^{j+2\lambda} =
1+(-1)^j =0$. This is so for any value of the twist $\theta^{\mu
\nu}$. Thus $Z_0$ cannot decay into two massless neutrinos of
helicity $\lambda$ for any value of $\theta^{\mu \nu}$ even if
lepton number is violated.

\section{Concluding Remarks}

We note that a relation of the form (\ref{twistmassivered2}) is
correct even for two identical massive particles. In that case,
(\ref{twistmassivered2}) is replaced by \be \mid j \lambda_1
\lambda_2 \ \widehat{p} \ J \mu \rangle^{S,A}_{\theta} =
\int_{SU(2)} d \mu (R) D^{J*}_{\mu \ \lambda_1 - \lambda_2}
(R)\frac{1\pm \tau_{\theta}}{2} \Delta_{\theta} (R) \mid q_1 \ j
\lambda_1 \rangle \mid q_2 \ j \lambda_2 \rangle.
\label{twistmassivered-1}\ee This reduces as before to \be \mid j
\lambda_1 \lambda_2 \ \widehat{p} \ J \mu \rangle^{S,A}_{\theta}
=e^{- {\frac{i}{2}}q_1 \wedge q_2} F^{-1}_{\theta} \mid j\lambda_1
\lambda_2 \ \widehat{p} \ J \mu \rangle^{S,A} .\ee It follows that
if ${\bar{\cal P}}^{\uparrow}_{+}$-invariance for $\theta^{\mu \nu}
=0$ forbids the decay of a spin $J$ particle into two identical spin
$j$ particles, then ${\bar{\cal P}}^{\uparrow}_{+}$-invariance for
$\theta^{\mu \nu} \neq 0$ also forbids it.

It is easy to show in a similar manner that if a decay into two
{\it{non}}-identical particles is forbidden by ${\bar{\cal
P}}^{\uparrow}_{+}$-invariance for $\theta^{\mu \nu} =0$, it remains
forbidden by ${\bar{\cal P}}^{\uparrow}_{+}$-invariance for
$\theta^{\mu \nu} \neq 0$.

Yang's result and those of this paper require two basic assumptions:
(a) the $S$-operator $S$ is invariant under ${\bar{\cal
P}}^{\uparrow}_{+}$, and (b) if $\psi_B (\psi_F )$ has a possibly
twisted Bose(Fermi) symmetry, $ S \psi_B ( S \psi_F )$ has the same
symmetry.

But not all QFT's on the Moyal plane share these properties. There
is in particular an approach to gauge theories with matter
\cite{gauge} which for non-abelian gauge groups gives Lorentz
non-invariant $S$-operators violating the Pauli principle. This
violation of Lorentz invariance by $S$ comes from the non-locality
of QFT's on the Moyal plane.

The standard model can be deformed along the lines of this approach.
The fate of the process $Z_0 \rightarrow 2 \gamma$ in this deformed
model is yet to be studied.

\vspace*{10mm} \noindent

{\bf Acknowledgments}

\vspace*{5mm}

A.P.B. thanks H. S. Mani for bringing Yang's paper to his attention
and for emphasizing the significance of the process $Z_0 \rightarrow
2 \gamma$. The work of A.P.B. is supported in part by US Department
of Energy under grant number DE-FG02-85ER40231. S.G.J. is supported
by the International Cooperation Research Program of the Ministry of
Science and Technology of Korea.

\end{document}